\documentclass[english,prl,a4paper,twocolumn,superscriptaddress]{revtex4}
\usepackage{graphicx}
\usepackage{amssymb}
\usepackage{babel}

\newcommand{\be}[0]{\begin{equation}}
\newcommand{\ee}[0]{\end{equation}}

\begin{document}

\title{Meta-nematic, smectic and crystalline phases of dipolar fermions in an optical lattice}

\author{Jorge Quintanilla}

\affiliation{ISIS spallation facility, STFC Rutherford Appleton Laboratory, Harwell
Science and Innovation Campus, Didcot, OX11 0QX, U.K.}


\author{Sam T. Carr}

\affiliation{School of Physics and Astronomy, University of Birmingham, Birmingham B152TT, U.K.}


\author{Joseph J. Betouras}

\affiliation{University of St. Andrews \& SUPA, North Haugh, St. Andrews KY16 9SS, U.K.}


\begin{abstract}
It has been suggested that some strongly correlated matter might be understood qualitatively in terms of liquid crystalline phases intervening between the Fermi gas and the Wigner crystal or Mott insulator. We propose a tunable realisation of this soft quantum matter physics in an ultra-cold gas. It uses optical lattices and dipolar interactions to realise a particularly simple model. Our analysis reveals a rich phase diagram featuring a meta-nematic transition where the Fermi liquid changes dimensionality; a smectic phase (stripes); and a crystalline, `checkerboard' phase.
\end{abstract}

\maketitle

A picture of strong correlations \cite{1998-Kivelson-Fradkin-Emery} unfolds as follows: as the strength of correlation increases, the Fermi gas condenses into a more correlated Fermi liquid. At this phase transition, the Fermi surface may change volume or even \cite{1999-Zverev-Baldo,2006-Quintanilla-Schofield} alter its topology. Then, the first `electronic liquid crystal' state forms: the nematic Fermi liquid \cite{2001-Oganesyan-Kivelson-Fradkin}, accessed through a symmetry-breaking deformation of the Fermi surface (a Pomeranchuk instability \cite{1958-Pomeranchuk}). As the strength of correlation increases further, a smectic phase develops. In this `striped' phase the Fermi liquid state is lost as electrons localise, but only in one direction. In the other direction the stripes behave as Luttinger liquids. Thus such liquid crystalline phases are intimately related to dimensional crossover phenomena (a subject of intense current interest, both in a condensed matter context \cite{2007-Giamarchi}
and in cold atomic gases \cite{2004-Ho-Cazalilla-Giamarchi}). Eventually, in the limit of very strong interactions, the particles localise completely, forming a Wigner crystal or a Mott insulator.

Experimental evidence for abrupt changes of Fermi surface volume or topology exists for heavy fermions \cite{heavyfermions123}. 
A nematic state is supported by transport measurements in YBa$_2$Cu$_3$O$_{6+y}$ \cite{2002-Ando-Segawa-Komiya-Lavrov} (with the transition rounded by lattice anisotropy).
There is evidence of nematic order in quantum Hall devices \cite{2002-Cooper-Lilly-Eisenstein-Pfeiffer}. A Pomeranchuk instability may explain `hidden' order in the heavy fermion URu$_2$Si$_2$ \cite{2006-Varma-Zhu} and the ruthenate Sr$_3$Ru$_2$O$_7$ \cite{2004-Grigera-et-al}. Smectic phases exist in manganites \cite{1996-Chen-Cheon} and cuprates \cite{1995-Tranquada-et-al}.   
In summary, there is evidence that elements of the scenario in Refs.~\cite{1998-Kivelson-Fradkin-Emery} resemble the physics of strong correlations. Yet in order to establish its general usefulness, a system that can be tuned from the Fermi gas all the way to the localised state and is amenable to theoretical treatment is necessary. 

In recent years it has become possible to realise strong correlations in highly-tunable cold atom experiments \cite{2007-Bloch-Dalibard-Zwerger}. Simple models, like the Hubbard model, can be realised precisely. Unfortunately even the 2D Hubbard model is very difficult to solve, even approximately. Experiments of that type must therefore be regarded as `quantum analogue simulations' \cite{2007-Campo-Capelle-Quintanilla-Hooley}. 
Here we propose an optical lattice set-up featuring dipolar fermions in an external field. The system consists of a 2D stack of chains, each of them containing
free fermions. In the absence of interactions the ground state is a non-interacting Fermi gas with
a nearly flat Fermi surface. Using an external field to produce a particular orientation of the dipoles relative to the lattice, we introduce a strictly inter-chain interaction as a perturbation, and address the stability of the 1D Fermi surface with respect to it. We argue that the system will feature a meta-nematic transition where the quasi-1D Fermi surface becomes fully 2D, competing with phase transitions into smectic and crystalline order.

%
\begin{figure}
\includegraphics[angle=90,width=0.9\columnwidth,keepaspectratio]{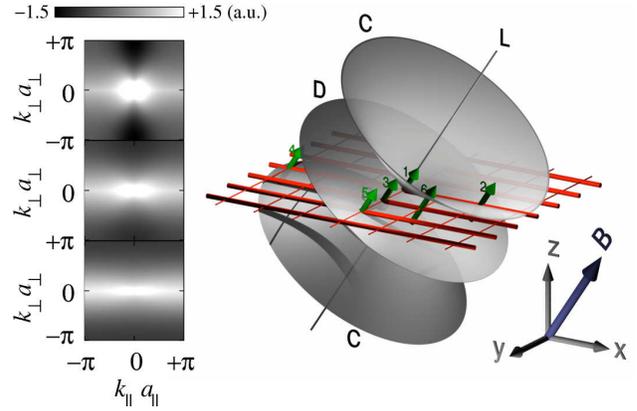} 
\caption{\label{fig:1}(color online) Right: Proposed experimental set-up. Dipolar atoms or molecules (1-6) are loaded on a 2D, anisotropic optical lattice in a strong external magnetic or electric field (B). The field is oriented so that all interactions between lattice sites are repulsive and there are no intra-chain interactions (see text). Left: 
\label{ftint}Interaction potential in reciprocal space for three values of the anisotropy parameter $\alpha = 0.5$ (top), $1.0$ and $2.0$ (bottom).
}
\end{figure}

%

The combination of optical lattices with dipolar interactions can be used to realise exotic Hamiltonians with novel states. This has been discussed extensively for bosons \cite{2006-Wang-Lukin-Demler,2007-Arguelles-Santos,2002-Goral-Santos-Lewenstein,2007-Astrakharchik-et-al,2007-Menotti-et-al,2007-Bochler-et-al}. Here we propose to use a \emph{fermionic} isotope with a large magnetic dipole moment, such as $^{53}\mbox{Cr}$ \cite{foot:Cr}. Alternatively, dipolar molecules \cite{2007-Sawyer-et-al,2008-Ni-et-al} or atoms cooled into a Rydberg state \cite{2000-Santos-et-al} may be used. Dipolar interactions between fermions in an external field are expected to display a range of interesting phenomena, including Fermi surface deformations (in 3D traps) \cite{2007-Miyakawa-Sogo-Pu}, exotic quantum Hall states (in rotating 2D traps) \cite{2007-Osterloh-Barberan-Lewenstein} and a `super Tonks-Girardeu regime' (in 1D traps, for either bosons or fermions) \cite{2007-Astrakharchik-Lozovik}. 

\emph{Proposed experimental setup.---} We propose to combine a polarizing external field (electric or magnetic, depending on whether we are exploiting the magnetic dipole of atoms or the electric dipole of molecules, for example) with an anisotropic, 2D optical lattice, as illustrated in Fig.~\ref{fig:1}. For sufficiently intense lasers, the lattice is in the tight-binding limit with one orbital per site \cite{2007-Bloch-Dalibard-Zwerger}. By allowing three different intensities for the three pairs of lasers we can make the system completely 2D and create ``chains'' along which hopping can occur, while keeping hopping perpendicular to the chains much smaller. Bonds along these two directions are represented by the thick and thin lines in Fig.~\ref{fig:1}, respectively. Tuning the effective wavelengths of the two in-plane lasers provides independent control of the lattice anisotropy 
\(
  \alpha \equiv a_{\|}/a_{\perp}.
  \label{eq:anisotropy}
\) 
Isotropic scaling of the lattice controls the relative strength of interactions.

We shall suppress the interaction between atoms or molecules that are on the same chain (for the sake of brevity, in what follows we assume the case of atoms in a magnetic field  without loss of generality). To this end we exploit the dependence of the dipole-dipole interaction, 
\(
  V\left(\mathbf{R}\right)
  =
  d^{2}[1-3\cos^{2}\theta]/\left|\mathbf{R}\right|^{3},
  \label{interaction}
\)
on the angle $\theta$ between the vector giving the relative positions
of the two dipoles, $\mathbf{R}$, and an external field, $\mathbf{B}$ strong enough to fully polarise all the atoms. We have represented this depedence schematically in Fig.~\ref{fig:1} by the line L, disc D and cone C around atom number 1. In these directions the interaction with another atom is maximally attrative [$\theta=0$], maximally repulsive  [$\theta=\pi/2$] and null [$\theta=\arccos\left(1/\sqrt{3}\right)\approx54.736^{\mbox{o}}$], respectively. In the proposed arrangement (see figure) the applied field is at precisely this latter `magic angle' to the chains and perpendicular to the inter-chain bond direction. Thus atom 1 does not interact with other atoms on the same chain, such as 2 and 4. Since on-site interactions are forbidden by Pauli's exclusion principle, in this setup there are no intra-chain interactions \footnote{Given the finite amplitude of the lasers, the width $\sigma$ of the on-site wave function is non-zero, leading to a small intra-chain interaction between atoms on different sites. Nevertheless as the interaction changes sign at the nodal suface C this is a very small effect, particularly for $\alpha \gtrsim 1$.}. Moreover inter-chain interactions are always repulsive and maximum in the direction of perpendicular hopping. The strongest repulsion corresponds to the closest sites on the two adjacent chains: $V\left(\mathbf{R}\right)=d^{2}/a_{\perp}^{3}$ (e.g. atom 3). Interactions with other sites on the two adjacent chains (e.g. 6) can also be made comparatively weak (see below).

\emph{Model.---} The single-particle Hamiltonian is 
\(
\hat{H}_{\mbox{hop}}=
-\sum_{i,l}\left(t_{\|}\hat{c}_{i,l}^{\dagger}\hat{c}_{i+1,l}+
t_{\perp}\hat{c}_{i,l}^{\dagger}\hat{c}_{i,l+1}
+\mbox{H.c.}\right)
,
\label{eq:H_hop}
\)
where $\hat{c}_{i,l}^{\dagger}$ creates a fermion on the $i^{\mbox{th}}$
site of the $l^{\mbox{th}}$ chain, $t_{\|}$ is the intra-chaing hopping amplitude and $t_{\perp}\ll t_{\|}$ is the inter-chain hopping. 
Defining the operator
creating a fermion with wave vector $\mathbf{k}=\left(k_{\|},k_{\perp}\right)$
by $\hat{c}_{\mathbf{k}}^{\dagger}=\sum_{j,l}\frac{1}{\sqrt{\Omega}}e^{-i\left(k_{\|}j+k_{\perp}l\right)}\hat{c}_{l,j}^{\dagger}$
we obtain $\hat{H}_{\mbox{hop}}=\sum_{\mathbf{k}}\epsilon_{\mathbf{k}}\hat{c}_{\mathbf{k}}^{\dagger}\hat{c}_{\mathbf{k}},$
which has an almost flat Fermi surface given by the zeroes of the `bare' dispersion relation, 
\(
\epsilon_{\mathbf{k}}=-2t_{\|}\cos\left(k_{\|}\right)
-2t_{\perp}\cos\left(k_{\perp}\right)-\mu
\label{eq:epsk}
\)
($\mu$ is the chemical potential).

The dipolar interaction has to be evaluated at the lattice sites. In the configuration discussed above, for relative coordinates $x\equiv a_{\|}i$ and $y\equiv a_{\perp}l$ (in units of the lattice constans in the parallel and perpendicular directions) it gives 
\(
	V_{i,l} 
	\propto 
	y^2/
	\left[
		x^2
		+y^2
	\right]^{5/2
	}.
\)
Fig.~\ref{ftint} shows the Fourier transform of this interaction, which depends strongly on the anisotropy ratio $\alpha$. For $\alpha \gtrsim 2$ it is well approximated by $V(k_{\|},k_{\perp})\approx 2V \cos(k_{\perp})$ i.e. nearest neighbour only interaction.
First we restrict ourselves to this limit. The interaction part of the Hamiltonian is thus
 \( 
 \hat{H}_{\mbox{int}}
  =
  V\sum_{i,l}
  \hat{c}_{i,l}^{\dagger}\hat{c}_{i,l+1}^{\dagger}\hat{c}_{i,l+1}\hat{c}_{i,l}
\),
and the full form is \footnote{The Hamiltonian in Eq.~(\ref{eq:H}) ignores the usual trapping potential. As is well-established theoretically \cite{2002-Batrouni-et-al} and experimentally \cite{2006-Folling-et-al} its effect is to `blur' the phase boundaries as different regions of the system go into different phases. The  relationship between translationally-invariant and trapped systems has been discussed in detail in Ref.~\cite{2007-Campo-Capelle-Quintanilla-Hooley}.}
\begin{equation}
  \hat{H}=\hat{H}_{\mbox{hop}}+\hat{H}_{\mbox{int}},
  \label{eq:H}
\end{equation}
The three parameters controlling our model are $\mu/t_{\|}$,  $t_{\perp}/t_{\|}$,  and $V/t_{\|}$. Below we discuss possible ground states.


\emph{Meta-nematic phase transition.---} We start by evaluating the stability of the Fermi surface shape. We use as a trial ground state a Slater determinant of plane waves, $\left|\Psi\right\rangle =\prod_{\mathbf{k}}\left[\left(1-N_{\mathbf{k}}\right)+N_{\mathbf{k}}\hat{c}_{\mathbf{k}}^{\dagger}\right]\left|0\right\rangle,$
 determining the occupation numbers $N_{\bf k}=0,1$ by requiring that the momentum distribution minimizes $\langle \Psi \left| \hat{H} \right| \Psi \rangle$. Such restricted Hartree-Fock mean field theory is similar to those used to study Pomeranchuk \cite{2004-Khavkine-Chung-Oganesyan-Kee,2006-Quintanilla-Schofield} and topological \cite{2006-Quintanilla-Schofield} Fermi surface shape instabilities. 

\begin{figure}
\includegraphics[width=0.85\columnwidth]{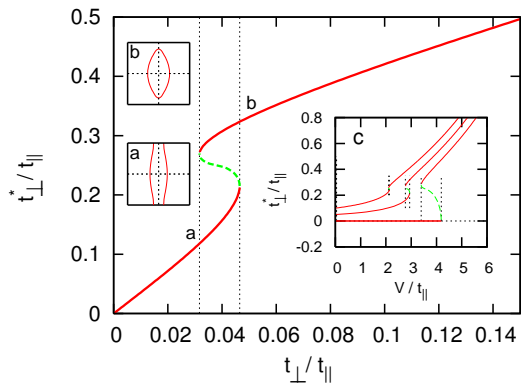}
\caption{\label{fig:3}(color online) Dependence of the renormalised transverse hopping $t_{\perp}^*$ on its bare value $t_{\perp}$ for $V=3t_{\|}$ and $\mu = -1.5
t_{\|}$. The solid (dashed) lines correspond to solutions to the self-consistency
equation (\ref{eq:tc*}) that minimize (maximize) the energy. Insets: Fermi surface (a) just to the left of the bifurcation region and (b) just to the right, as indicated; and (c) dependence of the solutions on $V$ for $t_{\perp}=0$ (rightmost curve), $0.05t_{\|}$
and $0.1t_{\|}$ (leftmost curve).}
\end{figure}

\begin{figure}
\includegraphics[width=0.82\columnwidth]{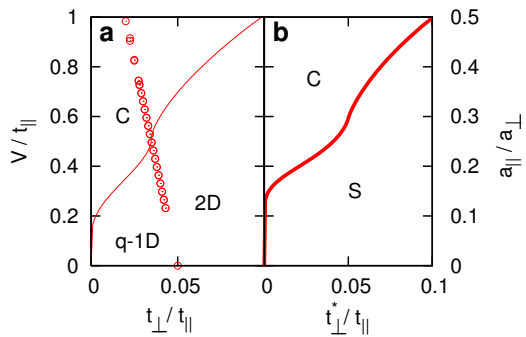}
\caption{\label{fig:Vcrit}\label{fig:alphacrit} (color online)
(a) Phase diagram of the Hamiltonian of Eq.~(\ref{eq:H}) for $\mu = \pm 1.9 t_{\|}$. The circles track the two bifurcation points (`a' and `b' in Fig.~\ref{fig:3}) of the first-order meta-nematic quantum phase transition between the quasi-1D and 2D phases. The solid line marks the line of quantum critical points separating these Fermi liquid states from the crystalline state (C). (b) Critical value of the lattice anisotropy $a_\|/a_\perp$ for the dominant instability to be towards the crystalline (C) or smectic phases (S) as a function of the relative strength of the renormalised perpendicular hopping, $t_\perp^*$, for $\mu^* = \pm 1.9 t_{\|}$.}
\end{figure}

The momentum distribution $N_{\bf k}$ corresponds to a non-interacting Fermi gas with a renormalized dispersion relation
\(
  \epsilon_{\mathbf{k}}^*=-2t_{\|}\cos\left(k_{\|}\right)
  -2t_{\perp}^*\cos\left(k_{\perp}\right)-\mu^*.
  \label{eq:epskr}
\)
The structure of $\hat{H}_{\mbox{int}}$ is such that only the
perpendicular hopping changes. It is given by the self-consistency
equation
\begin{eqnarray}
  t_{\perp}^{*} 
  & = & 
  t_{\perp}
  +
  \frac{V}{\Omega}\sum_{\mathbf{k}}\cos\left(k_{\perp}\right)N_{\mathbf{k}}.
  \label{eq:tc*}
\end{eqnarray}
Numerical solutions are shown in Fig.~\ref{fig:3}. As the bare
inter-chain hopping $t_{\perp}$ is increased, its renormalised value,
$t_{\perp}^*$, initially increases linearly but then has two bifurcation points,
between which lies a first-order jump. The corresponding phase
diagram is shown in Fig.~\ref{fig:Vcrit} (note that for Fig.~\ref{fig:3} we chose a very large value of $V/t_{\|}$, for clarity; for smaller values, the results are qualitatively the same, but the jump of $t_{\perp}^*$ is much smaller). The order parameter of this phase
transition is the amount of delocalisation in the perpendicular direction,
$\psi \equiv \langle \hat{c}_{i,l}^{\dagger}\hat{c}_{i,l+1}+\mbox{H.c.}\rangle$. We refer to
the jump of $\psi$ as we vary $t_{\perp}$ as a meta-nematic transition in analogy with
meta-magnetism (where the magnetisation jumps under an applied magnetic field).
As $V \to 0$, the meta-nematic transition becomes more and more weakly first order and requires a larger value of $t_{\perp}$. At $V=0$ there is no longer a first-order transition, but the phenomenon survives at $t_{\perp}=t_{\|}+\mu/2$ as a `two-and-half' order Lifshitz transition \cite{lifshitz} (while remaining frist order for any $V>0$). 

This quasi-1D to 2D transition induced by \emph{inter}-chain interactions is in some sense the opposite of confinement \cite{1994-Clarke-Strong-Anderson} (a quasi-1D to 1D transition induced by \emph{intra}-chain interactions \cite{2007-Ledowski-Kopietz}). A similar phenomenon is believed to occur in stacks of integer quantum Hall systems \cite{Betouras-Chalker}, where the
chiral Luttinger liquids on the edges couple
together, creating a 2D Fermi surface (the chiral Fermi liquid). In our
cold-atom set-up, the meta-nematic transition results from enhanced scattering when the potential reaches the singularities at the edges of the 1D bands. This is a density of states effect and hence we expect it to be robust to quantum fluctuations present for large values of $V/t_{\|}$ and not taken into account by our mean field theory. It could be induced by changing the intensity of one of
the lasers to tune $t_{\perp}$, and detected by direct imaging of the Fermi surface
\cite{2005-Kohl-et-al}. 

\emph{Crystallisation.---} The meta-nematic transition is not the only one possible in the sysem described by Eq.~(\ref{eq:H}).  Interchain backscattering can lead to `charge density
wave' (CDW) instabilities at low temperatures \cite{Klemm}.  We probe the potential CDW instability by examining the Fourier transform of the dynamic susceptibility, $X(\mathbf{k},t)=i\left\langle \Psi \left| T \rho(\mathbf{k},t)\rho^{\dagger}(\mathbf{k},0)\right|\Psi\right\rangle$. 
Here $\rho(\mathbf{k})=\sum_{\bf q} c_{\bf q}^{\dagger}c_{{\bf q}-{\bf k}}$ is the Fourier transform of the local occupation number, and $\rho({\bf k},t)$ is its Heisenberg representation. The `noninteracting' susceptibility (Lindhard function) is given by
\(
X_0 (\mathbf{k},\omega) 
= 
\int \frac  {d^2 \mathbf{q}}
            {(2\pi)^2}
      \frac {N_{\mathbf{q}} - N_{\mathbf{q}+\mathbf{k}}}
            {\omega - \epsilon^*_\mathbf{q} + \epsilon^*_{\mathbf{q}+\mathbf{k}}}
\)
in terms of the renormalised dispersion relation $\epsilon_{\bf k}^*$. 
Treating the interaction within the random phase approximation (RPA) gives 
$ X (\mathbf{k},\omega) = X_0 (\mathbf{k},\omega)/[1+2V\cos(k_{\perp}) X_0 (\mathbf{k},\omega)] $.
  An instability at wave vector $\mathbf{k}$ occurs if the static  component diverges, 
  \(
    X(k_{\|},k_\perp,\omega=0) \to \infty.
  \) 

For $t_{\perp}^* \ll t_{\|}$, $X_0 (\mathbf{k},\omega)$ is strongly peaked at $(2k_F,\pi)$ due to the strong nesting of the quasi-1D Fermi surface. Thus the system is unstable to a CDW of that periodicity at a critical coupling $V$ given the Stoner criterion
\(
    1=2VX_0(2k_F,\pi,\omega=0).
   \label{eq:rpa}
\)
In the limit $t_\perp^* \to 0$, the Fermi surface is perfectly nested and the peak in $X_0$ becomes a logarithmic divergence, implying $V \to 0$. More generally one has to evaluate $X_0(2k_F,\pi,\omega=0)$ to obtain $V$ {\it via} the Stoner criterion. The results are plotted in Fig.\ref{fig:Vcrit}~(a).

\emph{Smectic phase}.---  Within the nearest-neighbor approximation for the interaction we have been using so far,
the Fourier transform of the interaction potential $V(\mathbf{k})$ is independent of
$k_\|$, hence the dominant CDW instability is always at the peak of the Lindhard
function, i.e. $k_\|=2k_F$.  However, we now consider the full structure of the dipole
interaction.  In particular, as seen in Fig.~\ref{ftint}, when the ratio of lattice spacings $\alpha$ is not large, $V(\mathbf{k})$ acquires a large dependence on $k_\|$.
Hence so long as $t_\perp^* \ne 0$ (i.e. $X_0(k_\|=2k_F)$ is finite), then there is a level of anisotropy of the lattice where the leading instability is at $(0,\pi)$ and not $(2k_F,\pi)$.  

The $(0,\pi)$ instability is still a form of CDW - however as it breaks lattice symmetry in one direction only, it has smectic order.  Fig.\ref{fig:alphacrit}~(b) shows which instability takes place first as the overall strength of the interaction increases. Smectic order is favored when the fermions can lower their energy by crowding every other chain, paying a penalty in kinetic energy but taking advantage of the absence of intra-chain interactions. Note that the strong-coupling limit ground state is always a density wave. 

In the limit $t_{\perp} \to 0$, the system becomes 1D and one can employ bosonization. The bosonized Hamiltonian features backscattering terms responsible for the crystallisation at arbitrarily small coupling \cite{Klemm}. If we artifically turn off these backscattering terms we find a second instability characterised by a softening of the holon dispersion relation. The critical coupling for this second instability coincides with the divergence of the RPA susceptibility at $(0,\pi)$. Such accuracy of RPA is a result of a special feature of Hamiltonian (\ref{eq:H}), namely that interactions and single-particle dispersion take place in perpendicular directions in the limit $t_{\perp} \to 0$. Indeed as a result the lowest-order vertex corrections vanish identically in that limit. As $t_{\perp}$ grows such corrections will become increasingly important. For $t_{\perp} \sim t_{\|}$ strong correlation effects will modify the phase diagram, at least quantititatively. Describing these effects is beyond the scope of the present calculation.


\emph{In summary,} we have described a way to combine dipolar fermions in an external field with an optical lattice to realise a model featuring an array of chains with strictly inter-chain interactions. We have shown that the model has very rich physics, featuring competition between itineracy and localisation and between quasi one- and 2D behaviour. The possibility to realise meta-nematic, smectic, and crystallisation transitions in a regime where the system can be described using Hartree-Fock and RPA, together with the ability to introduce stronger correlations continuosuly into the system (for example, by tuning $t_{\perp}$ towards larger values) makes the proposed experiment a testbed for scenarios of correlated behaviour. 



We thank John Chalker for a stimulating suggestion; Andy Schofield, Mike Gunn and Eduardo Fradkin for useful comments. JQ gratefully acknowledges an Atlas Research Fellowship awarded by CCLRC (now STFC) in association with St. Catherine's College, Oxford. STC was funded by EPSRC grant GLGL RRAH 11382. 
JJB thanks the KITP, UCSB for hospitality when part of the work was done and was funded by Grand No. NSF PHY05-51164.

\end{document}